\documentclass[a4paper,12pt]{article}

\usepackage[margin=1in]{geometry} 

\usepackage{setspace} 

\usepackage{cmap}
\usepackage[T1]{fontenc}
\usepackage[american]{babel}
\usepackage{csquotes}
\usepackage{newtxtext,newtxmath}  
\usepackage{standalone} 
\usepackage{caption} 
\usepackage{subcaption} 
\usepackage{float} 
\usepackage{enumitem} 
\usepackage{tabularx} 
\usepackage{xcolor} 
\usepackage{soul}

\usepackage[
  backend=biber,
  style=apa,
  natbib=true,
  annotation=false,
]{biblatex}
\usepackage[hidelinks]{hyperref}

\title{Collective Benefits and Efficiency in Coupled Echo State Networks}
\author{Polyphony Bruna$^1$$^\dagger$ \and
        Rick Dale$^2$ \and
        Michael J. Spivey$^1$ \and
        Christopher T. Kello$^1$}

\date{
	$^1$University of California, Merced \\ \texttt{\{pbruna, spivey, ckello\}@ucmerced.edu}\\%
	$^2$University of California, Los Angeles \\ \texttt{rdale@ucla.edu}\\[2ex]
    $^\dagger$Corresponding author
}

\begin{document}
\maketitle
	
\begin{abstract}
Simple aggregation of independent responses ("wisdom of crowds") generally improves collective performance through statistical facilitation, but interaction among members can sometimes hinder group outcomes. However, under certain circumstances, coordination studies show that interacting groups can outperform their individual members and nominal groups, and interaction can also be more efficient than independence. We present a reservoir computing model to investigate how interaction can enhance joint performance in terms of accuracy and efficiency. Echo state networks were trained independently to identify Japanese speakers from vowel recordings and tested jointly with and without coupled feedback. Model pairs reproduced the dyadic advantage reported by Bahrami et al. (\citeyear{bahrami2010optimally}) with interaction via pooled-feedback, but not when independent responses were aggregated together. Model dynamics revealed that interaction was beneficial whereas aggregated outputs were prone to diverging from runaway feedback over time. In terms of efficiency, larger groups of smaller interacting models outperformed smaller groups of larger models while equating for total amount of group computation. Our results offer a formal framework for explaining when and why interaction enhances collective intelligence.

    
\end{abstract}


\section{Introduction}
	
Understanding how groups can exceed individuals on cognitive tasks is a central question of collective intelligence. One way that groups may be collectively intelligent is by exhibiting greater cognitive capacities than their individual members. For example, murmurations of starlings attain an extended sensory range compared to individual birds, allowing flocks to better respond to threats in their environment \citep{cavagna2010}. However, the cost of interaction between agents may exceed the relative performance gain of adding members. This interaction cost may be offset, however, if interacting groups can perform tasks more efficiently than the aggregation of individual efforts. For example, cetaceans may reduce the energetic cost of swimming when traveling in groups by exploiting hydrodynamics effects of nearby swimmers \citep{spina2024effect}.

Interaction in human group problem-solving has been shown to yield both performance gains and efficiencies. Interacting dyads can achieve greater outcomes than either of their best individual performances \citep{bahrami2010optimally, berger2026human}, and groups can also speed up decision-making by minimizing independent cognitive exploration \citep{dale2013dyadiccooperation, dale2014familiarity, almaatouq2021task, dreyer2025comparing}. At the same time, interaction may also suppress the benefits associated with independent exploration \citep{rotter1969group, janis2008groupthink}. Classic findings in group performance show that interaction can impair collective outcomes relative to equivalent non-interacting groups, such as by reducing the number of unique ideas generated during category recall \citep{szary2015memory, weldon1997collective}. Yet other approaches suggest that independent aggregation, without interaction at all, may yield the best statistical outcome \citep{surowiecki2005wisdom}. There are open questions about the underlying mechanisms driving these mixed observations. One approach to this puzzle is to investigate the underlying dynamics that interaction introduces to these collective decisions and the conditions under which such dynamics may facilitate performance.

When humans interact they can exhibit interpersonal synchronization across multiple biomechanical and behavioral dimensions, including gait, posture, physiology, gaze, head movement, and speech \citep{felsberg2021gaitsychronization, shockley2003mutual, yi2025respiratory, richardson2005looking, trujillo2023dynamic, brennan1996lexical, branigan2007syntactic, pardo2006phonetic, abney2014complexity, hamilton2026behavioral}. Although the functional role of synchrony is debated, it has been argued to support joint action, prosociality, mutual understanding, and the formation of linguistic conventions \citep{louwerse2012behavior, dale2013selforganization,  dasilva2025and, hove2009s, stolk2016conceptual, hawkins2023partners}. Consistent with this view, interpersonal synchrony has been linked to successful group problem-solving across multiple tasks \citep{abney2021cooperation, dideriksen2023quantifying}.

By contrast, aggregating independent guesses has been shown to produce a collective response that is more accurate than any individual guess, what is sometimes called the "wisdom of crowds" \citep{surowiecki2005wisdom}. This effect leverages a general benefit of statistical facilitation in certain problem spaces and was first demonstrated by pooling guesses of the weight of an ox \citep{galton1907voxpopuli}. Empirical illustrations of this statistical effect in general knowledge domains show that just a few pooled guesses from independent individuals can improve precision \citep{ariely2000effects}. These effects of statistical independence are sufficiently robust that some research has shown they can hold across multiple guesses from a single person \citep{vul2008measuring}. Nevertheless, the potential for bias within underlying distributions of individuals can raise concerns about how to achieve optimal outcomes \citep{centola2017socialinfluence} and the effect is still subject to factors such as relative level of knowledge across a population \citep{bang2017making}.

Although the general advantage of the wisdom of crowds is well-known, and cited as a canonical example of collective intelligence, it does not shed light on the cognitive role of human interaction. In the present study, we investigate the benefits of interaction in collective intelligence on both overall performance and on computational efficiency by developing a novel computational model of interaction dynamics in collective intelligence. We first simulate the finding that interacting dyads demonstrate a collective advantage in a joint perceptual decision-making task \citep{bahrami2010optimally} to shed light on how interaction can improve performance beyond the baseline benefit of statistical facilitation. We then investigate if collective performance yields efficiency gains when controlling for total computational cost.

\subsection{Performance Gains in Interacting Dyads}

Bahrami et al. (\citeyear{bahrami2010optimally}) provide us with a paradigm and experimental results to test our model of collective performance gains. They showed pairs of participants sets of Gabor patches (sinusoidal grating patterns with a Gaussian blur) in two consecutive intervals. Participants were tasked with identifying which interval, the first or the second, contained a contrast oddball. Individual responses were collected on each trial. When responses differed, participants conferred with each other to reach a joint decision. Individual and joint decisions across a range of contrast differences were fit with psychometric curves representing the perceptual sensitivity of each member and the dyad.

The study showed that some dyads' discriminability exceeded the perceptual sensitivity of their best member. Furthermore, the authors showed that this dyadic advantage was modulated by differences in individual member discriminability: Dyads composed of members with approximately similar perceptual sensitivities exhibited the greatest dyadic gain, and this gain diminished as the disparity between individual ability increased. The authors argued that this inverse linear relationship is consistent with the hypothesis that dyads communicated their level of confidence in their guesses to produce joint decisions \citep{fusaroli2012coming, bang2014does}.

There is an open question about the process by which dyads shared their confidences during joint decision-making. On the one hand, communication may only serve to integrate guesses after individual decision-making has terminated. In this case, the dyadic advantage would reflect the same baseline statistical facilitation also leveraged by wisdom of crowds. On the other hand, individual decision-making may continue during communication, and interaction between the decision processes themselves may produce an emergent benefit beyond what is attained by statistical facilitation. These two theoretical alternatives are illustrated in Figure~\ref{fig:model_pred}. In the present study, we investigate these alternatives by modeling dyads performing a classification task and testing if interaction or aggregation better accounts for the Bahrami et al. results. The question is whether or not interaction may benefit collective intelligence by enabling group performance gains above those that are explainable by statistical facilitation.

\subsection{Computational Efficiency in Interacting Collectives}

In addition to performance gains, interaction may also benefit collective intelligence by enabling groups to accomplish a task more efficiently. For example, recent findings showed that the force output of weaver ants increased superlinearly with group size such that force per ant nearly doubled when they coordinated as groups to pull and fold leaves into nests \citep{stewardson2025superefficient}. Alternatively, in humans, "social loafing," where effort per individual decreases with increasing group size, may grant collective efficiency if group performance is maintained or even increases despite the drop in individual effort. This is exactly what was found in a simulated real-world crisis mapping scenario: As group size increased, collective performance increased, effort per individual decreased, and collaboration across members increased, suggesting that interaction can distribute effort across human teams \citep{mao2016experimental}.

At the same time, while some studies show that the cost of interaction, as measured by response time, can outweigh benefits to accuracy \citep{straub2023cost, almaatouq2024effects}, other studies show that interaction can help groups respond more quickly and more accurately. For instance, one study found that group interaction led to more accurate decisions in monetary policymaking compared to individual decisions \textit{without} an increase in response time \citep{blinder2005two}. Speed advantages without loss in accuracy have also been reported in groups in mathematical reasoning tasks, "20-questions" -style games, general trivia, and crossword puzzles \citep{dale2013dyadiccooperation, dale2014familiarity, taylor1952twenty, laughlin2006groups}.

More recently, Hsieh et al. (\citeyear{hsieh2020new}) examined if dyads more efficiently process information in a contrast oddball task similar to Bahrami et al. (\citeyear{bahrami2010optimally}). They replicated the dyadic advantage and further showed that correct dyad responses were actually \textit{faster} and more frequent than would be expected by independent decision-makers. These findings suggest that dyadic interaction can result in a gain in both processing speed and perceptual sensitivity. We use the modeling framework developed herein to test for collective efficiency gains by measuring if performance improves with group size while controlling for the total amount of computational resources expended. Specifically, we test whether or not collective interaction among larger groups of computationally "cheaper" individuals leads to better performance than smaller groups of more "expensive" individuals, where computational resources are equal between comparison groups.

\subsection{Modeling Collective Dynamics using Ensemble Reservoir Computing}

Our approach to testing different hypotheses about collective performance is to model the ensemble dynamics of recurrent neural networks trained to perform a benchmark classification task. Recurrent connections are notoriously difficult to train and analyze, so we chose echo state networks (ESNs), a kind of reservoir computer that has three key benefits for our purposes. First, recurrent connections in ESNs are fixed and a read-out layer is configured to map their dynamics onto targets, which avoids complications with training recurrent weights, learning rates, and other associated hyperparameters. Second, ESNs are computationally inexpensive which facilitates the analysis of model interactions over time in groups of agents that vary in size. Lastly, ESNs can be easily coupled in different ways to test how interaction versus aggregation affects the collective dynamics of decision-making. 

In our simulations, we reproduce the dyadic advantage observed in Bahrami et al. (\citeyear{bahrami2010optimally}) using coupled ESNs, and we show that larger populations of interacting agents can achieve greater performance even when matched on overall computational cost. We first present a novel dyadic ESN architecture, which may be tested either as an aggregate of individual processes (\textit{qua} wisdom of crowds) or as a coupled information-processing system (\textit{qua} dyadic interaction). We test this model on a benchmark classification task to evaluate collective performance with and without interaction. We show that interaction via pooled-feedback over the course of processing produces a dyadic advantage similar to the original study. This effect is diminished with independent (self-) feedback, showing that interaction yielded a benefit beyond statistical facilitation. We analyze the network decision trajectories and show that pooled-feedback stabilizes network dynamics over time in unstable agents, allowing coupled networks to exhibit a collective benefit when they process the task as an interacting dyad.

We then expand this model architecture to simulate performance in larger interacting collectives. We control the quantity of nodes and edges in the network to produce a range of fixed computational resources available for information processing. We compare performance on the same signal classification task across group size and computational budget. We show a performance advantage in distributed, multi-agent architectures compared to individual networks matched on computational resources. The relative advantage increases with budget size and group size, suggesting an avenue for scaling reservoir computers by leveraging collective dynamics in distributed, coupled systems.

\section{Reservoir Computing and Echo State Networks}

Reservoir computing is an approach to neural network modeling that leverages the dynamics of a fixed, recurrent network of artificial neurons to perform general-purpose computations. When the dynamical memory of the reservoir balances signal retention with signal separation over time, the projection of inputs onto the nonlinear reservoir may be used to learn a linear "read-out" function, analogous to how biological intelligences may flexibly exploit naturally-occurring intrinsic dynamics for cognition without requiring task-specific wiring \citep{maass2002realtimecomputing}. Because the read-out function is linear, it can be solved directly using a least mean squares algorithm instead of a learning algorithm like backpropagation that requires gradual weight updates over repeated input presentations. As a result, reservoir networks offer simple, dynamical models of processing that have been applied to modeling the brain \citep{enel2016reservoir}, chaotic time series forecasting \citep{jaeger2004harnessing, pathak2018model}, speech recognition \citep{verstraeten2006reservoir}, and many other classification and prediction tasks. We employ ESNs as a dynamical model of decision-making that can be readily extended to collective processing among two or more networks.

\subsection{Modeling Methods}

We trained pairs of ESNs to classify speech signals from the Japanese Vowels dataset \citep{kudo1999japanesevowels} that consists of 640 recordings of nine native, male Japanese speakers pronouncing the vowel cluster /ae/ repeatedly. We randomly sample 28 recordings for each of the nine speakers for training data, which were cepstrum-coded to produce 12 linear prediction coefficients (LPCs) that encode the auditory signal on each time step, 25.6 msecs per step, with frames shifted by 6.4 msecs. Twenty-four additional recordings per speaker were sampled for the testing set. The first 10 LPC time steps of each recording were $z$-score transformed (recordings less than 10 time steps in length were excluded) and used as sequential inputs to the reservoir. The speaker identification task was coded by nine read-out units, where each unit represented one of the nine speakers and target vectors were one-hot coded.

Each ESN was composed of three layers: An input layer, a recurrent reservoir network, and a read-out (output) layer. The 12 LPCs were represented as 12-dimensional input vectors per time step, \(u(t)\), and connections from input units into the reservoir were defined by an input weight matrix \(W^{in}\). Input nodes were randomly connected to reservoir nodes with a probability of \(p_{inlink}\), and input weights were randomly set to 1 or -1 with equal probability. Symmetric weights serve two primary functions. First, zero-mean input weights (paired with zero-mean inputs) prevent a systematic directional bias in how external inputs perturb the reservoir. Second, high variance input weights increase the effective rank of the reservoir state matrix. This reduces correlations among reservoir nodes and maintains diversity in the reservoir state trajectories, which improves the linear separability of the resulting reservoir features.

The recurrent network consisted of \(N_{res}\) nodes connected to other nodes in the reservoir, including themselves, with a probability of \(p_{reslink}\). Recurrent connection weights, \(W^{res}\), were drawn from a Gaussian distribution (\(\mu=0, \quad \sigma=1\)) rescaled so that the weight matrix had a spectral radius of \(sr=0.9\). A spectral radius slightly under one is often used to balanced signal retention with signal separation \citep{lukovsevivcius2009reservoir}. Reservoir dynamics, \(x(t)\), were updated according to Equation \ref{eq:x_t}:

\begin{equation}
\label{eq:x_t}
x(t+1) = \textnormal{tanh}(W^{in} u(t) + W^{res} x(t) + W^{fb} y(t))
\end{equation}

where tanh squashes net input between -1 and 1. The tanh activation function is highly sensitive around zero, where external perturbations are centered, yet introduces nonlinear saturation for larger inputs. This helps the reservoir preserve variance across inputs while generating rich, nonlinear representations.

Lastly, the nine read-out nodes were fully connected to the reservoir and their weights were determined using the least mean squares algorithm for the whole training set of input/read-out pairs, with Tikhonov regularization \citep{tikhonov1977solutions}. Read-out units were also connected back to the reservoir with probability \(p_{inlink}\), and their weights were set randomly to 1 or -1 with equal probability. During training, the reservoir received teacher forcing as substitute feedback, where feedback is estimated using the target read-out signal on each time step \citep{lukovsevicius2007trainedfeedbacks} on the assumption that the ESNs would converge on their targets after training. Gaussian noise with a standard deviation of \(\sigma_{teacher}\) was added to each unit of the teacher forced signal to simulate the variability that will occur during testing. Variability may come from inaccurate read-outs or possible sources of noise, and teacher forcing is important to make feedback dynamics robust to such variability. 

\begin{figure*}[t]
    \centering
    \begin{minipage}[t]{0.69\linewidth}
        \centering
        \begin{subfigure}[t]{\linewidth}
            \captionsetup{justification=raggedright, singlelinecheck=false, position=top}
            \caption{}
            \label{fig:model_arch}
            \includegraphics[width=0.49\linewidth]{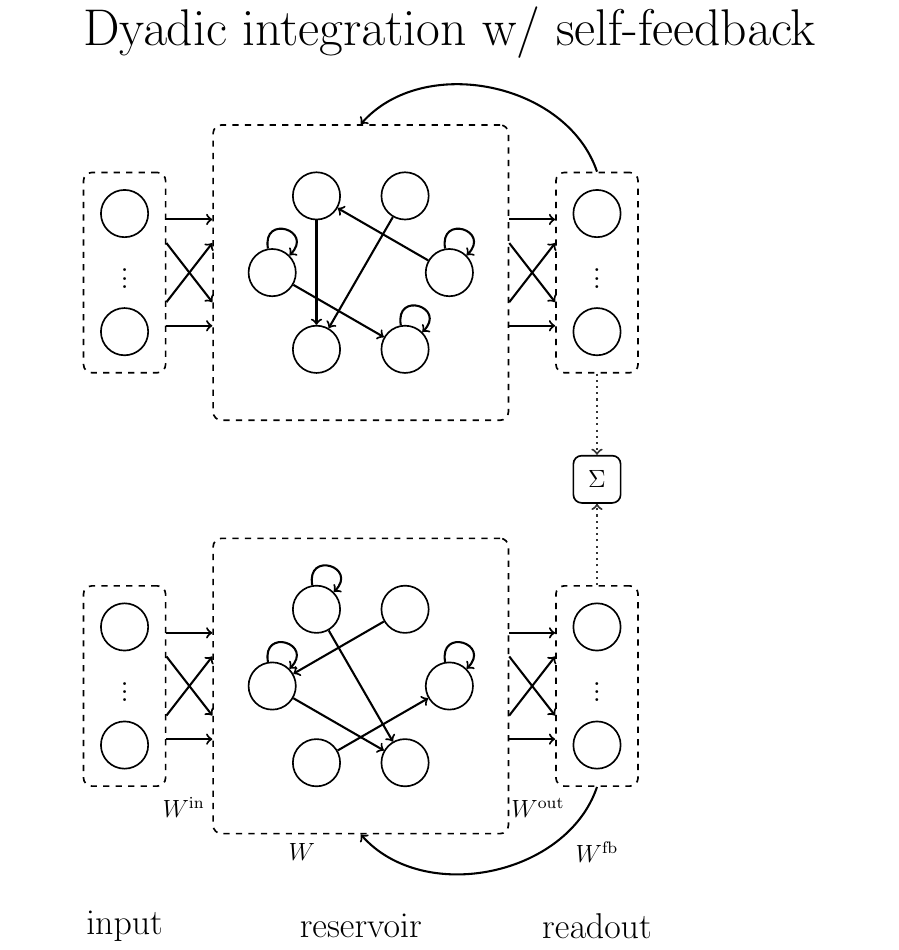}
            \includegraphics[width=0.49\linewidth]{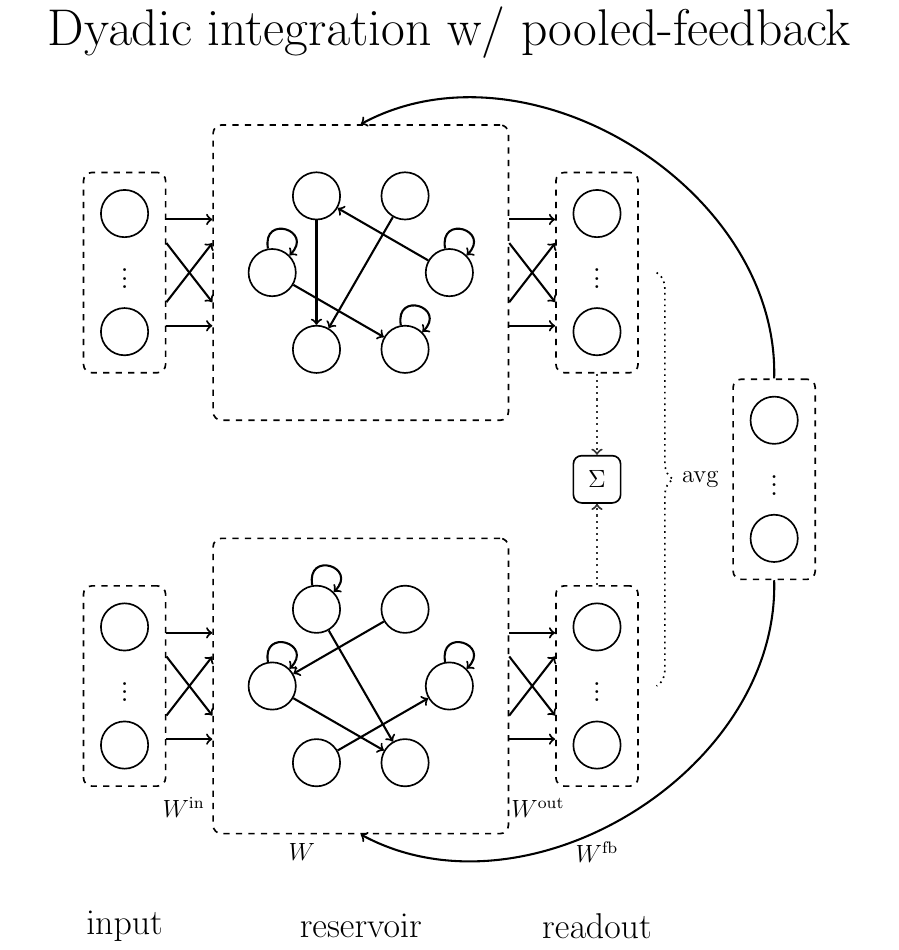}
        \end{subfigure}
    \end{minipage}
    \hfill
    \begin{minipage}[t]{0.3\linewidth}
        \centering
        \begin{subfigure}[t]{\linewidth}
            \captionsetup{justification=raggedright, singlelinecheck=false, position=top}
            \caption{}
            \label{fig:model_pred}
            \vspace{1cm}
            \includegraphics[width=\linewidth]{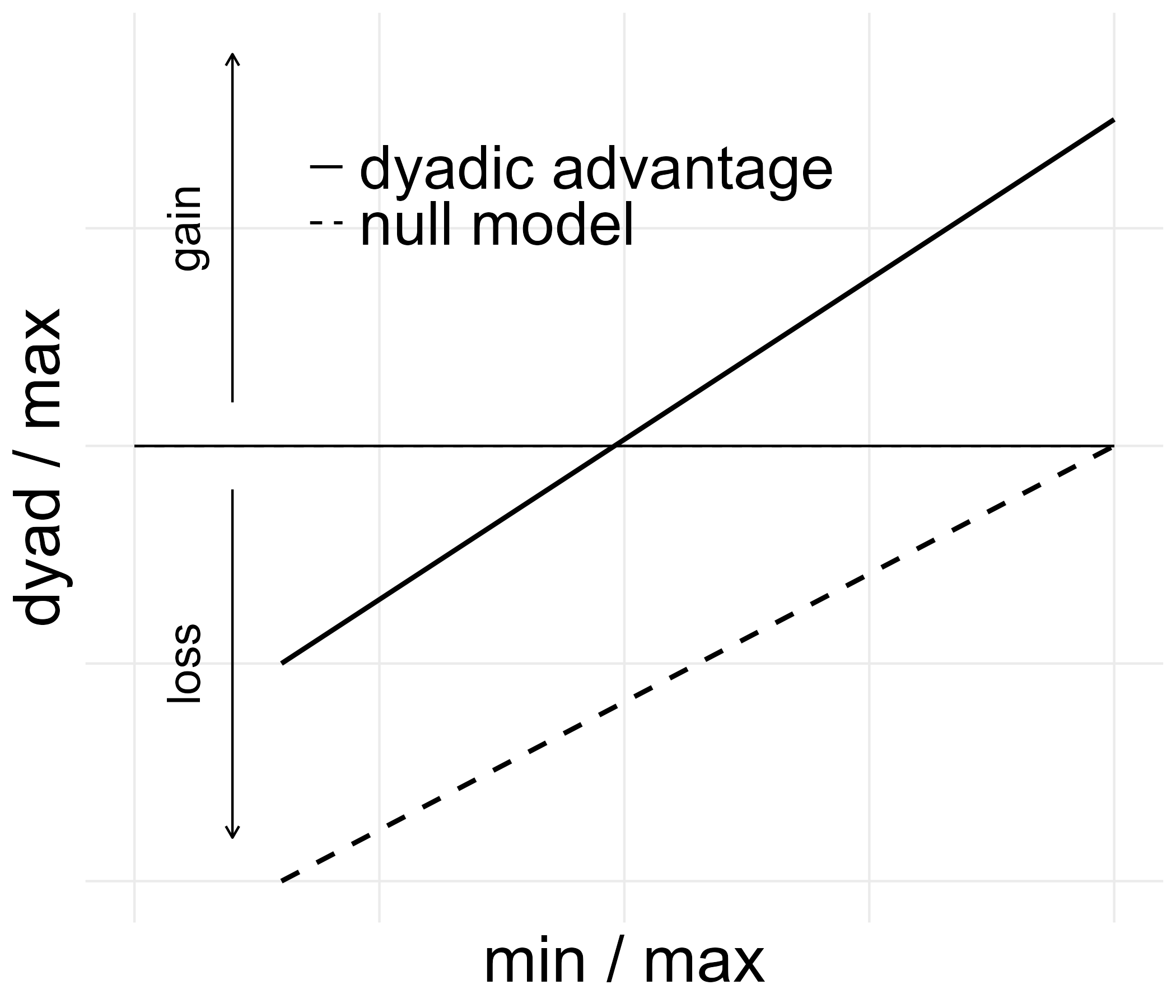}
        \end{subfigure}
    \end{minipage}

    \caption{\textbf{(a)} Schematic of the dyadic ESN model in each condition for one time step. In the self-feedback condition, read-out from independent ESNs is integrated node-wise across networks (shown by the summation box) as well as over time (not shown) in order to produce a collective response. In the pooled-feedback condition, ESNs are coupled through their feedback channels and each ESN receives the average of the two read-outs as feedback. \textbf{(b)} Theoretical predictions based on Bahrami et al. (\citeyear{bahrami2010optimally}). The null model is defined by dyadic performance equal to the average of its members. We aim to determine how much of the dyadic advantage is accounted for by statistical facilitation via self-feedback versus coupling via pooled-feedback.}
    \label{fig:model}
\end{figure*}

\section{Experiment 1: Dyadic Classification}

We use a standard reservoir computing classification task to examine performance gains in dyadic echo state networks. Networks either interact via coupled feedback channels, representing the communication of information pertaining to the relative likelihood of each possible decision, during information-processing or they receive their own feedback, completing decision-making independently. Dyadic performance is measured in both conditions by aggregating over network read-outs. Preliminary simulations not reported here indicated that classification performance on the Japanese Vowels dataset in a simple ESN does not depend on frame order. Consequently, these stimuli can be treated as equivalent to images, similar to the classification task used by Bahrami et al. (\citeyear{bahrami2010optimally}). Using these stimuli allows us to leverage echo state networks to test hypotheses about the dynamics that unfold over the time course of dyadic decision-making.

\subsection{Dyadic Architecture}

The dyadic ESN architecture is shown in Figure~\ref{fig:model_arch}. A joint decision was calculated from two trained ESNs by integrating the activation of their read-out layers over time and across networks. At the end of each signal, the integrated read-out node with the highest summed activation was taken as the decision of the dyadic network (Equation \ref{eq:argmax}).

\begin{equation}
\label{eq:argmax}
\text{Joint Decision} = \arg\max (\sum_{r=1}^{2} \sum_{t=1}^{10} y_{r,t})
\end{equation}

Networks were tested according to two conditions: Dyadic integration with self-feedback and dyadic integration with pooled-feedback. In the self-feedback condition, networks only received feedback from their own read-out layer, essentially completing the task independently and aggregating their independent outputs. In the pooled-feedback condition, networks received the average of the two read-out layers as feedback on each time step, allowing reservoir dynamics to influence each other during the processing of each signal. In both conditions, a joint decision was calculated on each test signal. This allowed us to measure collective performance from the model under both feedback conditions.

\subsection{Parameter Space}

\begin{figure*}[t]
    \centering
    \begin{minipage}[t]{0.7\linewidth}
        \centering
        \begin{subfigure}[t]{\linewidth}
            \captionsetup{justification=raggedright, singlelinecheck=false, position=top}
            \caption{}
            \label{fig:outcomes_psweep}
            \includegraphics[width=\linewidth]{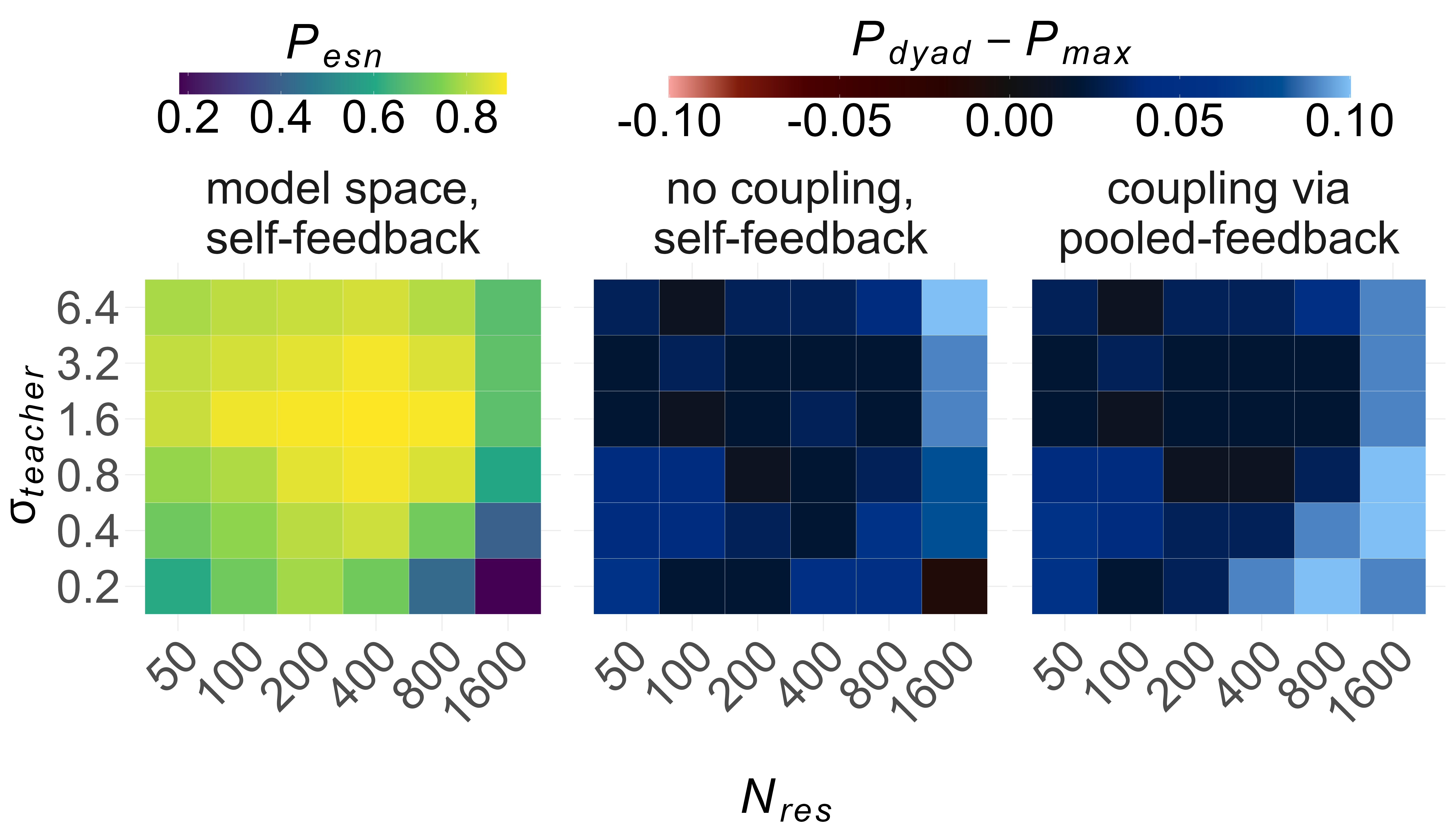}
        \end{subfigure}
        \begin{subfigure}[t]{\linewidth}
            \captionsetup{justification=raggedright, singlelinecheck=false, position=top}
            \caption{}
            \label{fig:outcomes_ratios}
            \includegraphics[width=\linewidth]{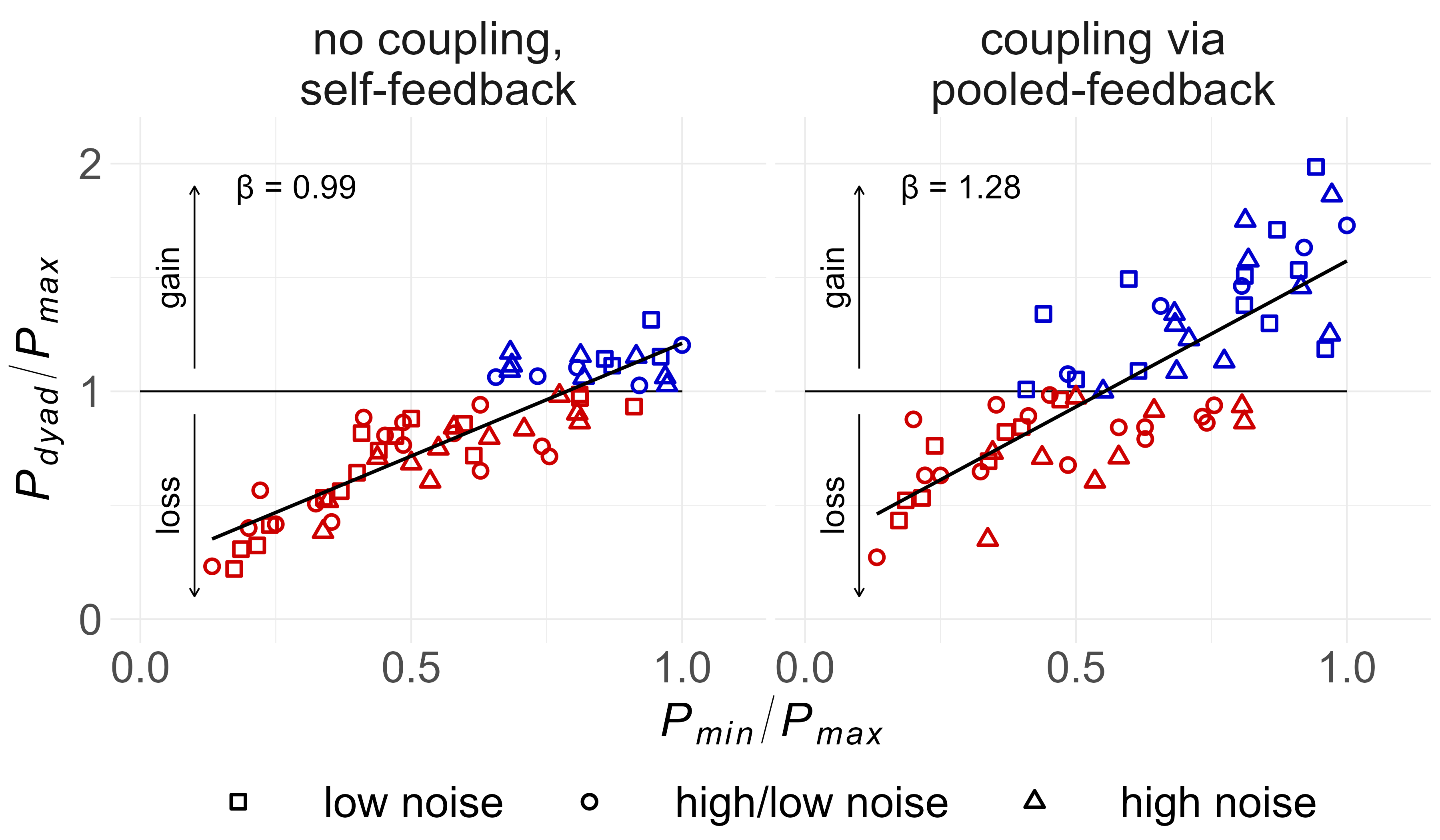}
        \end{subfigure}
    \end{minipage}

    \caption{
    \textbf{(a)} Performance as a function of reservoir nodes (\(N_{res}\)) and noise on teacher forcing (\(\sigma_{teacher}\)). \textbf{(b)} Collective benefit compared to member disparity across both conditions. Values above \(P_{dyad}/P_{max}\)=1 indicate the dyad outperforms its best member and values below \(P_{dyad}/P_{max}\)=1 indicate the dyad underperforms its best member.}
    \label{fig:outcomes}
\end{figure*}

Individual ESN performance on the speaker classification task varied along two key parameters defining the model space: Size of the reservoir (\(N_{res}\)) and noise added for teacher forcing (\(\sigma_{teacher}\)). Performance within this model space averaged over 20 simulations is shown in Figure~\ref{fig:outcomes_psweep}. Performance varied nonlinearly, with accuracy on the task reaching ceiling with moderate \(N_{res}\) and \(\sigma_{teacher}\) (unreported simulations showed a comparable trend without feedback). This pattern did not change substantially by varying the reservoir or input connectivity, so we held them constant at (\(p_{reslink}=0.1\) and \(p_{inlink}=0.1\)).

Figure~\ref{fig:outcomes_psweep} also illustrates dyadic performance (averaged over 10 simulations) for the best member of each dyad, for both conditions across the parameter values tested. Results show a small dyadic advantage over the best member for most values, and there was a a substantial dyadic advantage for coupled networks localized to the bottom-right corner of the figure, corresponding to a region where individual performance drops off. To investigate this difference in collective performance, we sampled ESNs randomly from \(N_{\mathrm{res}} \in \{800, 1600\}\) and \(\sigma_{\mathrm{teacher}} \in \{0.2, 0.4\}\) to generate heterogeneous dyads and test them against the results from Bahrami et al. (\citeyear{bahrami2010optimally}), shown in Figure~\ref{fig:outcomes_ratios}. Bahrami et al. add noise to their stimuli to induce performance variability across their participants. To match these experimental conditions, we added Gaussian noise to each input node: "High noise" (\(\sigma_{input}=2\) for both networks), "low noise" (\(\sigma_{input}=0.5\) for both networks), and "high/low noise" (one network received high noise and the other received low noise).

\subsection{Results}

We implemented the model and ran simulations using ReservoirPy \citep{reservoirpy2020}, and analyses were performed in the R environment. Results were averaged over 10 simulations for dyads, or 20 simulations for individual networks. Performance was measured as proportion correct on test signals.

\subsubsection{Coupled ESNs Produce a Dyadic Advantage}

Following Baharami et al. (\citeyear{bahrami2010optimally}), we examined the relationship between member disparity and collective benefit for 60 simulated dyads (20 per noise condition), where the ratio of worse over better performing member of each dyad corresponded to member disparity, and the ratio of dyadic over best individual performance corresponded to collective benefit. Figure~\ref{fig:outcomes_ratios} shows the relationship between member disparity and collective benefit under the self- and pooled-feedback conditions. Both conditions display an inverse relationship between collective benefit and member disparity, indicating that dyads with members that were matched on individual performance tended to perform better collectively. A mixed-effects linear regression with dyad as a random effect revealed a significant interaction between feedback condition and member disparity, indicating that the relationship between member disparity and collective benefit was stronger with pooled-feedback than self-feedback (\(\beta=0.29, \textnormal{p}=0.016\)). Estimated marginal means (EMMs) from this regression were used to test collective benefit at various levels of member disparity in each condition.

Consistent with results from Bahrami et al. (\citeyear{bahrami2010optimally}), pooled-feedback showed a significant collective benefit for dyads with matched members (\(\textnormal{EMM}=0.57, \textnormal{p}<0.0001\)) that persisted until the worse member performed roughly half as well as their partner, after which dyads began to underperform their best member due to the worse member dragging down performance. In contrast, the magnitude of this effect was reduced with self-feedback. In matched dyads, self-feedback showed a significant, but smaller, collective benefit (\(\textnormal{EMM}=0.21, \textnormal{p}<0.0001\)). In closely matched dyads (\(P_{min}/P_{max}=0.75\)), pooled-feedback still showed a significant improvement over the best member (\(\textnormal{EMM}=0.25, \textnormal{p}<0.0001\)), but self-feedback did not (\(\textnormal{EMM}=-0.037, \textnormal{p}=0.22\)).

Bahrami et al. (\citeyear{bahrami2010optimally}) found that dyadic performance approximated that of the best member when the ratio of the worse to better member was roughly 0.42, which is more closely fit by pooled-feedback (0.55) compared to self-feedback (0.79). Most dyads with self-feedback exhibited a collective detriment relative to their best member (72\%), compared to pooled-feedback (53\%) and the empirical results (32\%). Together, these findings indicate that the pooled-feedback condition offers a better qualitative fit to Bahrami et al., suggesting that the dyadic advantage depends on interaction between members during decision-making.

\begin{figure*}[t]
    \centering
    \begin{minipage}[t]{0.52\linewidth}
        \centering
        \begin{subfigure}[t]{\linewidth}
            \captionsetup{justification=raggedright, singlelinecheck=false, position=top}
            \caption{}
            \label{fig:network_dynamics}
            \includegraphics[width=\linewidth]{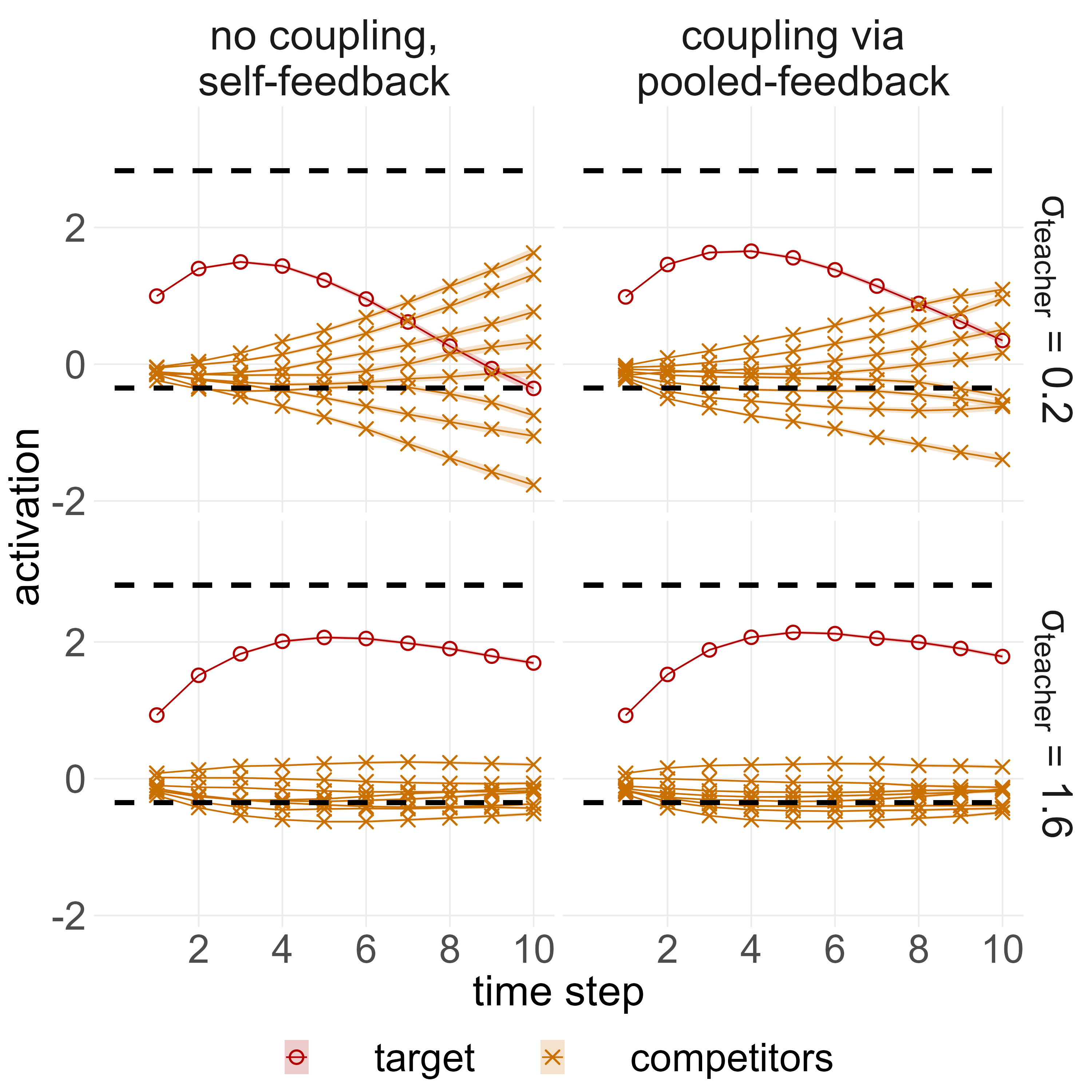}
        \end{subfigure}
    \end{minipage}
    \hfill
    \begin{minipage}[t]{0.45\linewidth}
        \centering
        \begin{subfigure}[t]{\linewidth}
            \captionsetup{justification=raggedright, singlelinecheck=false, position=top}
            \caption{}
            \label{fig:network_delta}
            \includegraphics[width=\linewidth]{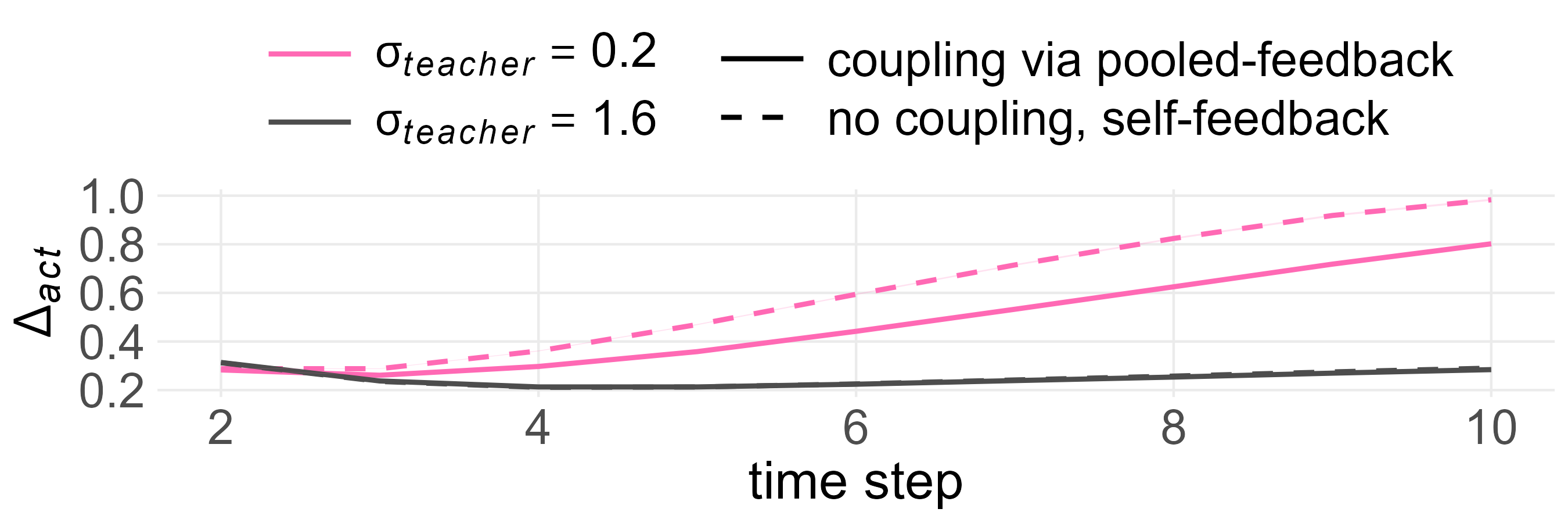}
        \end{subfigure}
    
        \vspace{-0.25cm}
    
        \begin{subfigure}[t]{\linewidth}
            \captionsetup{justification=raggedright, singlelinecheck=false, position=top}
            \caption{}
            \label{fig:network_noise}
            \includegraphics[width=\linewidth]{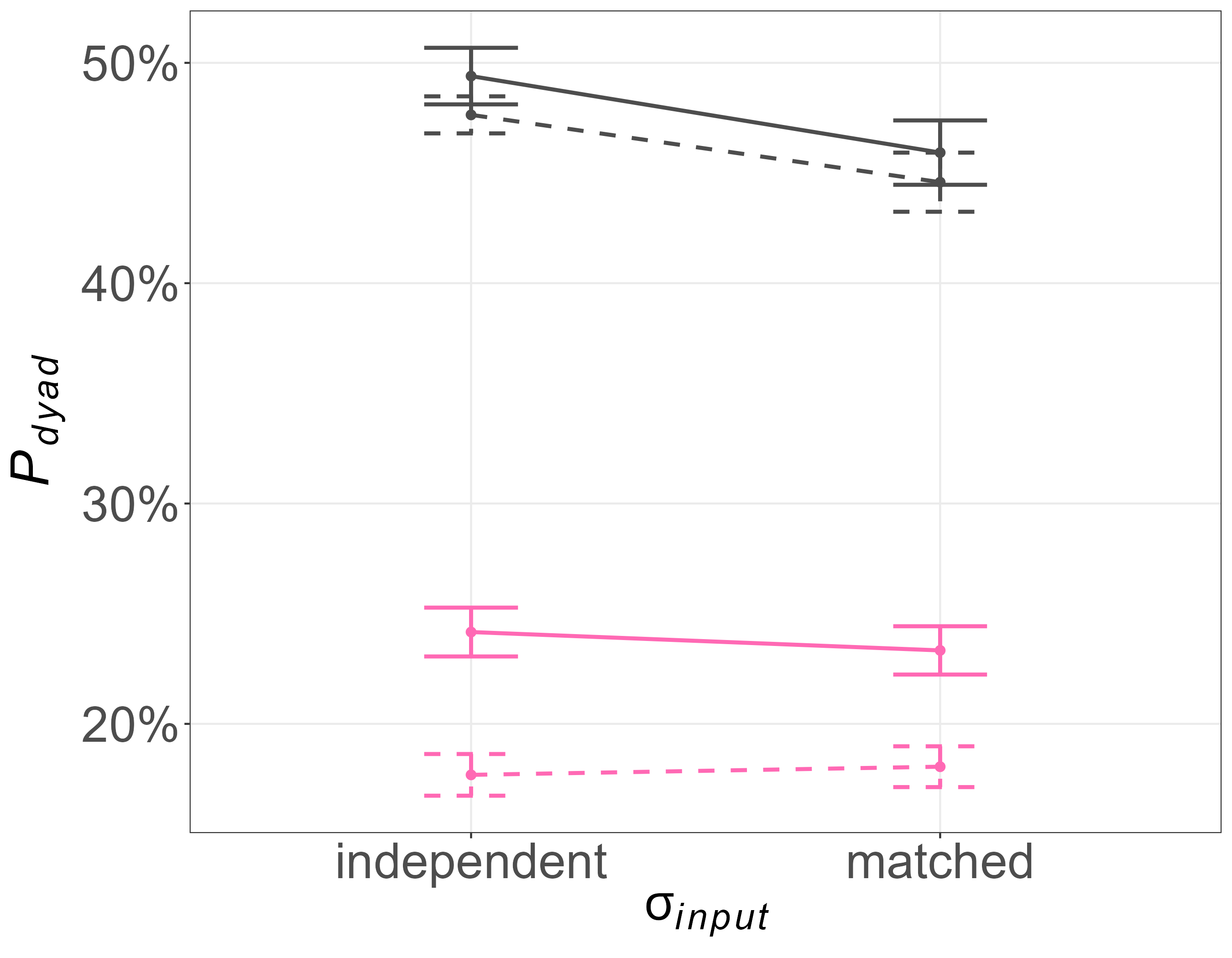}
        \end{subfigure}
    \end{minipage}

    \caption{
    \textbf{(a)} Time series of read-out node activation across both conditions for two levels of \(\sigma_{teacher}\). Networks are matched on reservoir size (\(N_{res}=800\)). Incorrect trajectories are averaged to preserve the rank-order of the decision. Dashed lines indicate the target activation value for the correct and incorrect class. \textbf{(b)} Absolute first-order temporal difference averaged over read-out nodes for each condition and two levels of \(\sigma_{teacher}\). Ribbons indicate standard error. \textbf{(c)} Performance with independent and matched input noise for each feedback condition and two levels of \(\sigma_{teacher}\), representing unstable (left) and stable (right) networks. Networks are matched on reservoir size (\(N_{res}=800\)). Bars indicate standard error.}
    \label{fig:network}
\end{figure*}

\subsubsection{Interaction Stabilizes Network Dynamics}

Next we compared the dynamics in the read-out layer over the time course of signal processing with self- and pooled-feedback to investigate how interaction alters the decision process of networks in the dyadic model. Figure~\ref{fig:network_dynamics} shows time series for each of the nine decisions averaged over signals. Time series were compiled such that the first trajectory (plotted with o's) corresponds to the correct response and the following trajectories (plotted with x's) indicate competing decisions ordered by rank.

With self-feedback, networks failed to retain the correct classification decisions until the end of a signal, demonstrated by trajectories that increasingly diverge from their target activation as signals persist. However, with pooled-feedback this divergence was mitigated. Networks maintained more stable decision trajectories over time, shown by a delay in how quickly, and to what extent, each trajectory diverges in the coupled networks. The first-order temporal difference (Equation~\ref{eq:tempdiff}) averaged over test signals was used to measure the dispersion of the trajectories over time in each condition (Figure~\ref{fig:network_delta}). A linear mixed-effects regression model with condition as a predictor of dispersion and random effects of node, signal, and time step confirmed that trajectories with pooled-feedback dispersed significantly less over time compared to self-feedback (\(\beta=-0.13, \textnormal{p}<0.0001\)).

\begin{equation}
\label{eq:tempdiff}
\Delta(t)_{act}=\frac{1}{9} \sum_{i=1}^{9} |y_{i}(t)-y_{i}(t-1)|
\end{equation}

These findings suggest that pooling the feedback of two networks over time stabilizes the coupled dynamics compared to integration of independent network outputs. The benefit of pooling vanished for networks trained with stronger teaching forcing to be more robust to the effects of variability feedback dynamics (\(\beta=-0.0015, \textnormal{p}=0.15\)), suggesting that a dyadic advantage may arise when agents are prone to runaway positive feedback in their decision-making process. Interaction may introduce a negative feedback dynamic that stabilizes individual behavior and enables dyads to outperform their best individuals.

\subsubsection{Correlated versus Independent Input Noise}

To simulate the experimental conditions of Bahrami et al. (\citeyear{bahrami2010optimally}), Gaussian noise was added to the inputs for each ESN that was sampled \textit{independently} of noise added to other ESN inputs. Therefore, pooling may have helped by averaging out these independent noise sources, or it may have stabilized intrinsic reservoir dynamics that differ across ESNs due to weight randomization and teacher forcing. We tested these two possibilities by comparing performance in self- and pooled-feedback conditions with independent input noise (per the original simulation) versus identical input noise for each ESN in a dyad (noise was always sampled independently across dyads). Input noise was set to \(\sigma_{input}=2\) to match the "high noise" condition from the first simulation, and the same parameter settings were used to compare unstable and stable network architectures.

Figure~\ref{fig:network_noise} compares dyadic performance with independent versus matched input noise for dyadic networks in both feedback conditions. Unstable reservoirs showed a benefit of pooled-feedback, consistent with prior results, and the benefit persisted in networks with matched input noise. This result indicates that pooling was beneficial at least in part because it stabilized reservoir dynamics. By contrast, stable reservoirs showed a weaker benefit of coupling and a relatively large effect of noise type, such that pooling was more beneficial with independent input noise. Therefore, pooling can help to average out independent noise for stable networks, but it seems that input noise cannot be averaged out when it is mixed with unstable dynamics.

\subsection{Summary}

We show that the dyadic advantage reported by Bahrami et al. (\citeyear{bahrami2010optimally}) can be reproduced by coupled ESNs. When individual networks are prone to instability interaction can stabilize their dynamics, yielding distributed improvements and a superadditive benefit in collective performance. Networks were trained on stimuli and targets with an average expected value of zero, allowing read-outs to represent confidence over classification targets (with the most confident target class taken as a response). Pooling read-outs to generate coupled feedback therefore provides a simple mechanistic implementation of confidence sharing, hypothesized to underlie the original effect \citep{bahrami2010optimally, bang2014does}. Importantly, the model shows how confidence sharing shapes joint dynamics and identifies the conditions under which it produces a collective advantage.

Nevertheless, the dyadic advantage requires two individuals to solve a shared problem, doubling the computational resources involved. One may ask if the performance gain persists when dyads are matched to individuals in total computational cost. This question is difficult to address in biological collectives, where adding members necessarily increases computational resources; however, our modeling framework allows computational cost to be controlled directly. In experiment two, we compare individual and ensemble ESNs of varying sizes while holding total computational cost constant. Because biological collectives generally coordinate through interaction rather than by aggregating independent processes, we test coupled ensembles to determine if distributed computation is more efficient even in the face of the added cost of coordination. Thus, we ask not only if "two heads are better than one," but if \textit{more} heads are better \textit{for the price of} one.

\section{Experiment 2: Efficiency of Group Interaction}

Dyads can outperform their individual members, through coupling or statistical facilitation, but it may be at the expense of increased cognitive work done or effort exerted by two individuals instead of just one. Studies have shown that dyads and groups can enjoy an efficiency advantage in terms of time on task \citep{dale2013dyadiccooperation, dale2014familiarity, almaatouq2021task}, but it is difficult to empirically measure efficiency in terms of effort or work. In our model, however, it is straightforward to measure work in terms of the amount of computation performed by an ESN. Given that the amount of processing is constant per time step, the computational cost of an ESN can be expressed in terms of its total number of units and connections. Furthermore, if we set aside the computational cost of interaction as small compared with the cost of each ESN, then we can quantify the computational cost of a group of ESNs simply as the sum of units and connections across their reservoirs.  

In Experiment 2, we investigate the efficiency of collective performance in our model by extending it from dyads to larger ensembles and measure performance across ensembles of different ESN numbers and sizes while holding computational cost constant across groups. This control afforded by reservoir computing allowed us to test the degree to which more interactions among smaller ESNs are more efficient than fewer interactions among larger ESNs. While we did not quantify the cost of interaction, we can estimate whether or not it matters by measuring performance across a range of ESN and group sizes to determine if increasing costs affect results.

\subsection{Method}

\subsubsection{Generalized Coupled Architecture}

The dyadic architecture from experiment one was generalized to support multi-agent ESNs of ensemble size \(E\). Networks were trained on the same classification task from experiment one and tested using pooled-feedback. Collective decisions were again calculated by integrating ESN read-outs over time and across networks, with the highest summed activation node taken as the collective decision (Equation \ref{eq:argmax_collective}). Read-outs were trained with teacher forcing on the reservoir at each value tested in Experiment 1. Collective decisions were generalized as

\begin{equation}
\label{eq:argmax_collective}
\text{Collective Decision} = \arg\max (\sum_{r=1}^{E} \sum_{t=1}^{10} y_{r,t})
\end{equation}

\subsubsection{Parameter Space}

Ensembles sizes of \(E \in \{1, 2, 4, 8, 16\}\) were tested with \(N_{\mathrm{res}} = 2^b \times 10,\quad b \in \{0,\ldots,7\}\) reservoir sizes. Internal reservoir connectivity, \(p_{reslink}\), was varied by reservoir size to maintain an average out degree (\(k_{\mathrm{out}}\)) of 10 (Equation \ref{eq:sparsity}). This control allowed us to hold number of units and number of connections constant for a given total computational cost. Moreover, reservoir connectivity is known to have minimal impact on ESN performance and is often fixed via out-degree instead of connection probability in larger reservoirs to allow for linear scaling properties \citep{lukosevicius2012}. Linear scaling allowed us to match ensembles for total number of units and connections; e.g., an individual ESN of \(E=1, N_{res}=20, p_{reslink}=0.5\) and a dyadic architecture of \(E=2, N_{res}=10, p_{reslink}=1\) both yield 20 total reservoir units and 200 total reservoir connections. \(N_{res}\) and \(p_{reslink}\) values were chosen such that reservoir connectivity ranged from fully saturated, \(p_{reslink}=1\), in the smallest reservoirs while the largest reservoirs did not breach the Erd\H{o}s--R\'{e}nyi connectivity threshold (Equation \ref{eq:er_thresh}).

\begin{equation}
\label{eq:sparsity}
p_{\mathrm{reslink}} = \frac{k_{\mathrm{out}}}{N_{\mathrm{res}}},
\qquad \text{where } k_{\mathrm{out}} = 10.
\end{equation}

\begin{equation}
\label{eq:er_thresh}
p_{\mathrm{reslink}} > \frac{\log N_{res}}{N_{res}}
\end{equation}

Reservoir weights \(W^{res}\) and input weights \(W^{in}\) were sampled from the same distributions as in experiment one, and input weights were randomly connected to the reservoir with a fixed probability of \(p_{inlink}=0.1\). Reservoir weights were fixed at initialization, and the reservoir matrix was rescaled to obey a spectral radius of \(sr=0.9\). In order to examine performance across various ensemble sizes, Gaussian noise was simulated for each input node (\(\sigma_{input}=2\)), and input noise was matched across all ESNs in each group.

\subsection{Results}

Model simulations were produced using ReservoirPy (Trouvain et al., 2020), and analyses were performed in the R environment. Results were averaged over 60 simulations for each ensemble architecture, 10 simulations per \(\sigma_{teacher}\) value, yielding sufficiently small standard errors per parameterization relative to the effect sizes reported below. The pattern of results does not change over \(\sigma_{teacher}\) values. Performance was measured as percent correct on test signals.

\subsubsection{Dyadic Advantage Extends to Larger Ensembles}

\begin{figure}[t]
    \centering
    \includegraphics[width=0.7\linewidth]{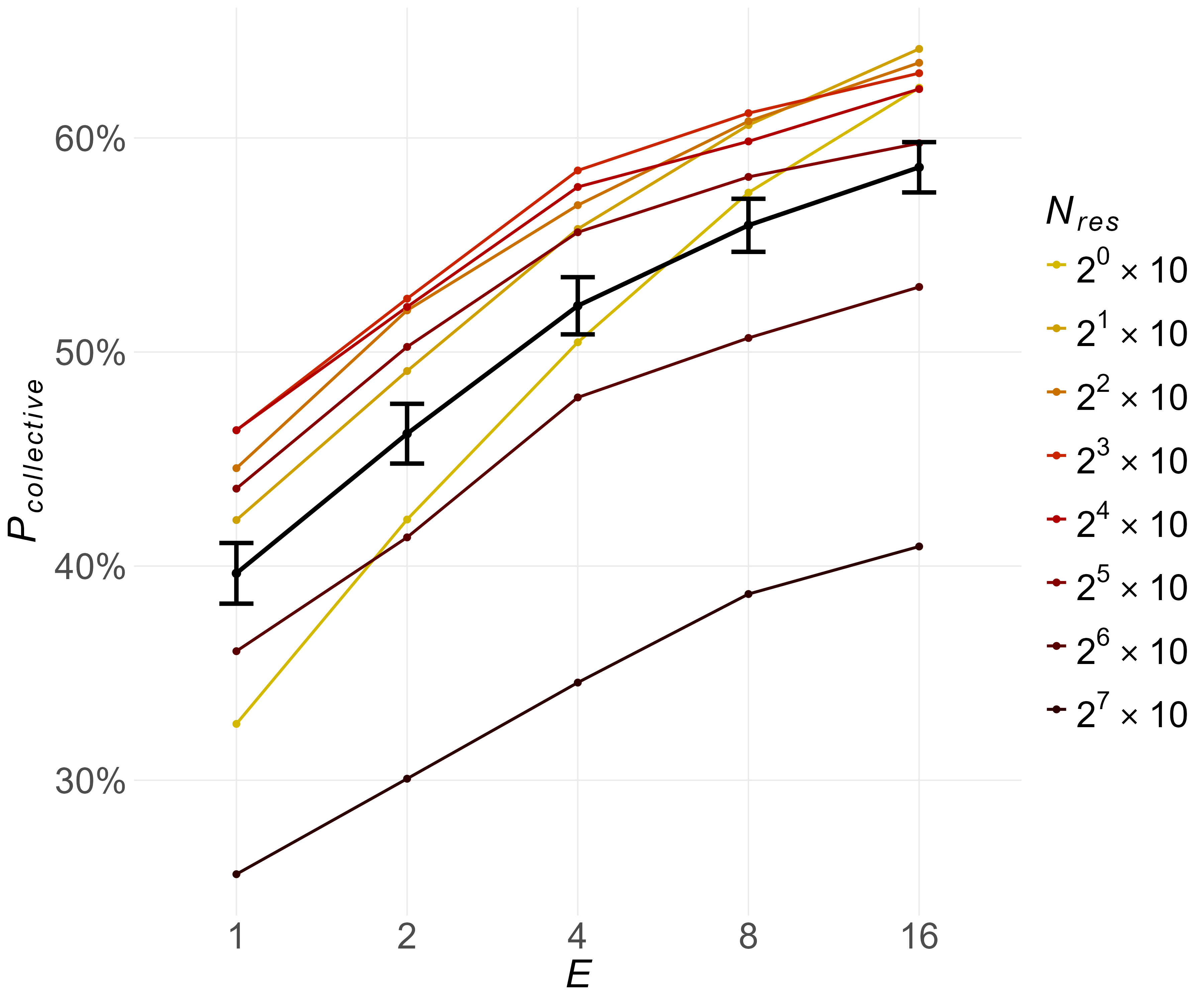}
    \captionsetup{justification=raggedright,singlelinecheck=false}
    \caption{Performance with increasing group sizes. Average performance is shown in black. Bars represent standard error over reservoir sizes. The effect of \(\sigma_{teacher}\) was consistent across reservoir sizes and is not shown.}
    \label{fig:ensemble}
\end{figure}

First, we looked at the overall effect of group size on performance. Figure~\ref{fig:ensemble} shows a monotonic increase in performance with more ESNs interacting, averaged over all tested reservoir sizes. Compared to an individual ESN, ensembles of 16 ESNs exhibited an improvement of almost 20 percentage points. Performance appears to begin to asymptote at 16 ESNs, although it is unclear how many ESNs can be added while performance gains continue to accrue. 

Figure~\ref{fig:ensemble} also shows that the rate of increase in performance differs for ensembles composed of different reservoir sizes. Ensembles composed of 10-unit reservoirs exhibit the strongest effect of group size. Single 10-unit reservoirs show some of the lowest performance, but ensembles of 16 show equivalent performance as larger networks. This result suggests that larger ensembles of small reservoirs can compete with the performance of larger networks for less total computational cost.

\subsubsection{Larger Ensembles of Smaller Networks Are More Efficient}

\begin{figure*}[t]
    \centering
    \begin{subfigure}[t]{0.48\linewidth}
        \centering
        \captionsetup{justification=raggedright, singlelinecheck=false, position=top}
        \caption{}
        \label{fig:efficiency_matched}
        \includegraphics[width=\linewidth]{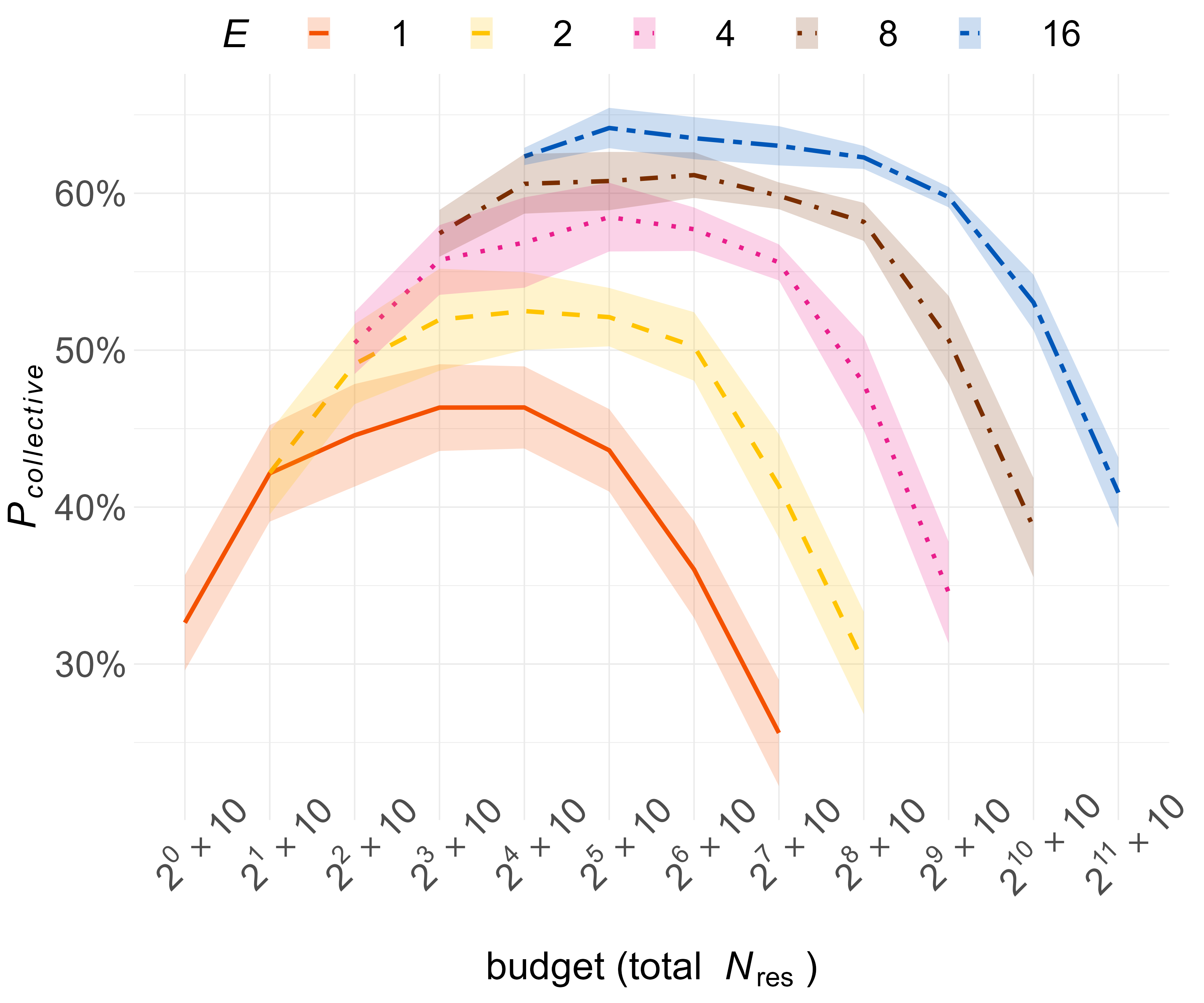}
    \end{subfigure}
    \hfill
    \begin{subfigure}[t]{0.48\linewidth}
        \centering
        \captionsetup{justification=raggedright, singlelinecheck=false, position=top}
        \caption{}
        \label{fig:efficiency_diff}
        \includegraphics[width=\linewidth]{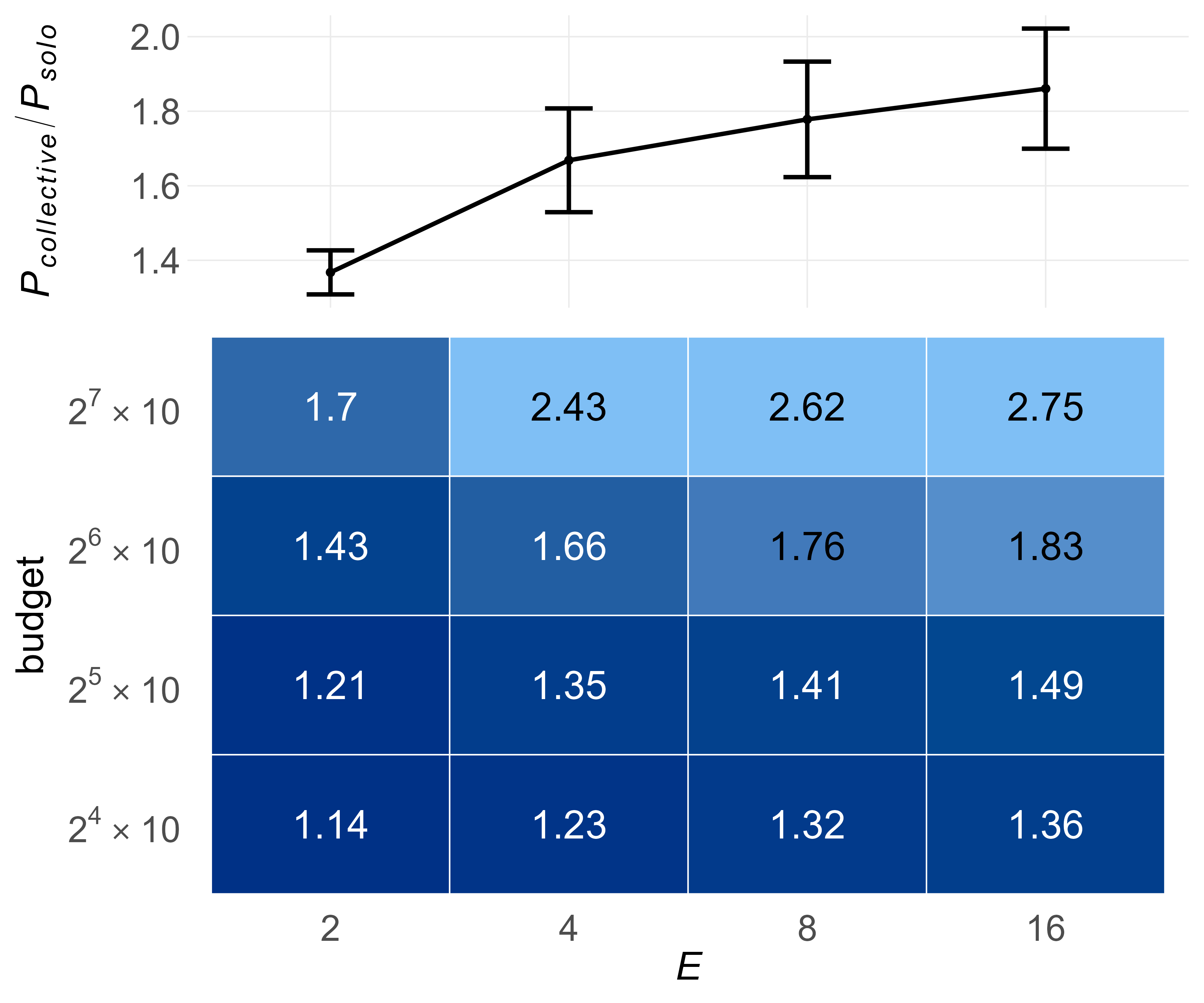}
    \end{subfigure}
    \captionsetup{justification=raggedright, singlelinecheck=false}
    \caption{
    \textbf{(a)} Performance per ensemble over increasing budget sizes. Ribbons represent standard error over \(\sigma_{teacher}\) values. \textbf{(b)} Relative performance compared to individual network baseline for ensembles matched on computational budget. Average is shown above in black; bars represent standard error over budget sizes.}
    \label{fig:efficiency}
\end{figure*}

Figure~\ref{fig:efficiency_matched} compares ensembles that are matched for total computational cost at each point along the x-axis, where computational cost is doubled at each step. For each given ensemble size, including individual ESNs, a non-monotonic performance curve was found such that increasing ESN size yielded better performance up to a point, and further increases caused declines in performance due to well-known limits of reservoir size \citep{jirak2023echo}. 

For our purposes, increasing ensemble size yielded better performance for every computational budget and ESN size tested, even when accounting for additional interaction costs--for computational costs that total twice or more than their minimum cost per ensemble size (10 units per ESN), the performance advantage for each increase in ensemble size holds true even if costs are doubled to account for interactions. Interaction costs are far less in our model because they only involve pooling and broadcasting read-out values, which scale linearly with ensemble size. Natural interactions may cost and scale differently and it is likely to depend on the kinds of agents and interactions.   

Finally, Figure~\ref{fig:efficiency_diff} quantifies the efficiency gains for each ensemble size by plotting the ratio between collective and individual ESN performance at matched budgets. The figure shows that, while both the mean and range of efficiency gains increased with ESN size, they both leveled off at a mean gain of 1.86 over individual ESNs for the maximum ensemble size of 16, about double the performance gain for dyads with the same computational cost. In summary, group performance scaled up with ensemble size more than ESN size; the best average performance we observed was for 16 ESNs with reservoirs of only 20 units each. While this scaling effect may be attributed to ESNs and how read-outs were pooled, our model nonetheless demonstrates how ensemble dynamics can scale up with ensemble size to support collective performance.

\subsection{Summary}

Our model allows systematic investigation into the collective benefit of interaction in larger groups by controlling underlying factors such as the computational budget of individuals. Our findings with groups are consistent with the benefits of pooled-feedback observed in dyadic networks in the first experiment. We also show that the relative gains from interaction depend on the group size and capacity of the networks. We observe that smaller, less capable networks in larger groups generally benefit more from pooling information compared to smaller groups of larger networks. These findings suggest that distributing computation among relatively simple agents can be advantageous not only in terms of collective outcomes but in terms of collective efficiency. Our results also suggest an avenue for scaling reservoir computers by leveraging distributed computation, which we discuss below.

The ability to systematically vary group size and computational capacity also offers a framework that can be readily extended to test factors in optimal team composition. Prior work on team composition suggests a number of factors impact group outcomes, including relative performance of individuals, group size, and group structure \citep{lamberson2012optimal}. These factors can be translated into testable predictions within our model and used to investigate potential underlying mechanisms systematically \citep{mcclelland2009place}. ESNs implement well defined and parameterized processes that can be precisely tuned for computational budget and individual training as well as leveraged across many tasks. This framework could thus contribute new mechanistic links between cognitive processes and collective outcomes \citep{scott2022cognition,smaldino2023problems}.

\section{General Discussion}

In this paper, we ask what dynamics enable dyads and groups to perform better when deliberating together versus independently and how interaction can be leveraged for efficient collective decision-making. We offer a novel model of joint decision-making using coupled ESNs that shows how dyads can outperform their best individual member. They can do so somewhat through statistical facilitation, but interaction through step-by-step pooling of read-outs yielded even better performance and also better accounted for the data. This benefit of coupling ESNs via interactions grew further still with larger ensemble sizes, in terms of absolute performance as well as computational efficiency. Our model shows how interactions among reservoir networks can shape their individual dynamics in the service of collective decision-making. Reservoir computing is general-purpose and our model could be used to investigate the role of dynamics in many different kinds of collective tasks and performances.

\subsection{Interaction and Crowd Wisdom}

In our model, pooling feedback offers a simple mechanism to fulfill the function of weighted confidence-sharing proposed by Bahrami et al. (\citeyear{bahrami2010optimally}). On each time step, read-out vectors encoded both a decision (most active node) and a degree of confidence in whether the vowel token came from each of the nine speakers--specifically, unit magnitudes encoded confidence because their expected values were zero when the reservoir provided no reduction in the uncertainty of speaker identification. As identification information becomes available in the reservoir, units move away from their zero baselines towards identification of one or more speakers over others. Therefore, a more confident ESN contributed larger read-out magnitudes compared with a less confident ESN, and larger magnitudes exerted more influence on collective decisions.

Weighted confidence-sharing does not specify how interaction alters decision-processing over time. Individual guesses could be combined after independent decision-making has terminated, which would predict that integrating decisions across independent networks would show the same dyadic advantage as coupled networks, or better. Instead, we found that pooling feedback step-by-step as inputs came in over time provided a better account of the data, which we take as evidence that the collective benefit in perceptual decision-making and similar tasks can emerge from interactions that shape both individual and collective processing dynamics.

Our model also shows how structured interaction can enhance the benefit of crowd wisdom. For example, Navajas and colleagues (\citeyear{navajas2018aggregatedknowledge}) found that small-group consensus can outperform simple aggregation of individual estimates when participants deliberate independently before interacting. Likewise, we observe that while aggregating independent network decisions yields a modest collective benefit, this benefit increases in both magnitude and scope when pooled-feedback is introduced. There is also evidence that variability among individual judgments is important for crowd wisdom \citep{barreralemarchand2024divergentopinions}. Likewise, it was important that we randomized and trained our ESNs individually so their dynamics would be distinct during testing, allowing them to make distinct contributions to interactions and their collective effects via pooled-feedback. Finally, network structure and response weighting can amplify or suppress crowd wisdom \citep{baten2025humblenetworks, centola2017socialinfluence}. Such effects could be simulated and tested with our model, and simulations of larger social networks under various coupling rules may help us understand the functional connections between individual and social dynamics.

The generality of our framework aside, we focused on a classic nine-choice speaker classification benchmark so we could investigate and analyze the model in depth. Bahrami et al. (\citeyear{bahrami2010optimally}) used a two-alternative visual discrimination task, so it would be useful to test if simulations more specific to their task that estimate visual discriminability from psychometric curves might provide additional insights into their results. It would also be useful to simulate other tasks and results that both do and do not confer a benefit to collective interaction.

\subsection{Collective Stabilization of Positive Feedback}

We found that the benefit of interaction over statistical facilitation only held for relatively unstable reservoirs that needed pooling to guide their runaway dynamics. Runaway dynamics came from positive feedback between each reservoir and its read-out units, and such dangers of positive feedback are common and well-known in neural and behavioral dynamics. In the case of decision-making, positive feedback can manifest as "overthinking," such as when a student revises a correct multiple-choice answer because deliberation continues after sufficient evidence has been reached. One can imagine that interaction with a peer might help the student avoid overthinking and stick with their original answer.

Positive feedback can also enable exploration and adaptation but unchecked it runs the risk of causing instability \citep{brandman2008feedback, may1972will}. For example, attempts to maintain a constant vocal volume often lead to unintentional amplification \citep{zollinger2011lombard}, while excessive positive feedback in brain circuits can produce seizures \citep{jirsa2014nature, huguenard2007thalamic}. Negative feedback stabilizes neural dynamics, supporting homeostasis, and disruptions in the balance of positive and negative feedback contribute to psychiatric disorders like schizophrenia \citep{molina2020memantine}. Neural computation is thought to emerge near this balance at the "edge of chaos" \citep{shew025criticality, kello2013critical}, a regime also exploited by reservoir computing to maximize both stability and flexibility \citep{bertschinger2004realtime}.

Our work demonstrates that balancing positive and negative feedback can emerge from the interplay of individual and collective dynamics. Agents like reservoir networks that are individually susceptible to positive feedback can achieve stable performance by leveraging negative feedback loops introduced through dyadic and group interaction. This finding extends the idea that nonlinear interactions across scales can transform the dynamical properties of individual components and overall behavior to collective systems \citep{kelty2023multifractal}. In particular, collective near-critical dynamics can emerge from interactions among agents whose individual dynamics are not themselves optimally tuned \citep{bessone2026emergent}. Thus our model highlights a functional mechanism available to groups with collective tasks and goals.

\subsection{Collective Efficiency}

Collectively intelligent systems raise a fundamental question about the relationship between the organization of computation and the resources required to perform it \citep{resendiz2021shrunken}. Collective advantages, such as the dyadic advantage reported by Bahrami et al. (\citeyear{bahrami2010optimally}), suggest the possibility that the benefit of interaction dynamics can allow collectives of relatively simple agents to outperform a more computationally costly system. If so, collective interactions may be an evolutionarily favorable strategy for collective intelligence. The challenge of evolving ever more complex \textit{organisms} may be overcome by instead evolving more complex \textit{interactions} among relatively simple agents. Our model demonstrates how more efficient decision-making can emerge from larger ensembles of small networks rather than smaller ensembles of larger networks.  

Our finding of collective efficiency is partly due to the lack of scaling in reservoir networks like ESNs. Their recurrent dynamics are nonetheless useful for modeling biological and behavioral dynamics \citep{schubert2021local}, as well as predicting chaotic time series \citep{jaeger2004harnessing, pathak2018model}. For instance, the synchronization of chaotic time series has been modeled using an ensemble reservoir approach similar to ours \citep{hu2022synchronizationreservoir}. However, the limited scalability of reservoirs has so far made them particularly favorable in machine learning applications where computational resources are limited, like edge computing \citep{kokalj2021reservoir}. Some previous studies have shown that ensembles of ESNs can perform better on some machine learning benchmarks \citep{giovannozzi2023ensemblereservoir}, and our findings on ensemble scaling suggest future directions for research into ensemble reservoir computing.

Most notably, our approach differs from previous efforts that use more complex topologies and means of collective processing \citep{qiao2016growing, viehweg2025systematic}. Our model is less costly in terms of computation and yet yields substantial performance improvements. The model also lends itself to further improvement through ESN pre-training to create so-called "mixtures of experts" \citep{jacobs1991adaptive, ghahramani2000variational}. This well-known approach to machine learning is highly effective across a wide range of applications, including large language models \citep{shazeer2017outrageously, fedus2022switch} and convolutional neural networks that aggregate network outputs with adaptive soft mixture-of-experts \citep{hoang2026cross}. It is also highly efficient because only small handfuls of experts may be needed to process input. In our model, further improvements may be gained by training low-cost ensembles of specialized ESNs. For example, negative correlation learning \citep{yao1999negativecorrlearning} is a technique in which neural networks are trained to produce negatively correlated outputs and it has been shown to yield better performance in ensemble reservoirs \citep{tino2011nclreservoir}. Along with mixtures of experts, this finding is consistent with evidence that crowd wisdom improves in maximally divergent crowds \citep{barreralemarchand2024divergentopinions}. Taken together, these studies along with ours motivate further research into collective performance and efficiency using interactions in reservoir ensembles.

\subsection{Conclusion}
Models of both natural and artificial collective intelligence have been gaining traction in recent years with advances in data collection and analysis \citep{mcmillen2024collective, centola2022network} as well as computational resources and techniques \citep{flint2026group, casadei2023artificial, kong2026messaging}. Ensemble reservoir computing offers a viable approach to modeling collective intelligence in general by leveraging the computational efficiency of reservoirs with tractable interactions to investigate the dynamics that enhance as well as limit or undermine collective performance. It also offers a point of intersection and potential cross-pollination among machine learning researchers and behavioral and social scientists interested in collective intelligence.

\section{Acknowledgments}

Scripts, data, and other materials are available at: \url{https://github.com/pjbruna/coupled-esns}. Part of this research was performed while Polyphony Bruna was visiting the Institute for Pure and Applied Mathematics, which is supported by the National Science Foundation (Grant No. DMS-1925919), and the Santa Fe Institute (Graduate Workshop in Computational Social Science; Advanced Graduate Workshop). This project also received support from the Institute for Humane Studies (Grant Nos. IHS019149 and IHS018920) and from a UC Merced Cognitive and Information Sciences Fellowship to Polyphony Bruna. AI was not used in the preparation of this project and manuscript.

\printbibliography

\end{document}